\documentclass[twocolumn]{aastex631}

\usepackage{graphicx}
\usepackage{xcolor}
\usepackage{soul}
\usepackage{color}

\newcommand{\delm}{$\Delta$mag}

\begin{document}

\title{The sensitivity of TESS to transiting planets in TOIs with close-in stellar companions}

\author[0000-0001-7746-5795]{Colin Littlefield}
\affiliation{Bay Area Environmental Research Institute, Moffett Field, CA 94035, USA}

\author[0000-0002-9903-9911]{Kathryn~V.~Lester}
\affil{Mount Holyoke College, South Hadley MA 01075, USA}

\author[0000-0002-5741-3047]{David~R.~Ciardi}
\affiliation{NASA Exoplanet Science Institute, Caltech/IPAC, Pasadena, CA 91125, USA}

\author[0000-0002-2532-2853]{Steve~B.~Howell}
\affil{NASA Ames Research Center, Moffett Field, CA 94035, USA}

\correspondingauthor{Colin Littlefield}
\email{littlefield@baeri.org}

% ----------------------------------------------------------------------
\begin{abstract}

    High resolution imaging with optical speckle interferometry has revealed that many transiting exoplanet host stars possess close-in stellar companions. The objective of this study is to quantify how the presence of these companions impacts the ability of TESS to detect the transits of small planets. We accomplish this by examining 2052 TESS Objects of Interest (TOIs) that appear to be single-star systems based on speckle interferometric observations as well as 188 TOIs in unresolved ($< 1.2\arcsec$) stellar binaries. For each planet, we take its transit signal-to-noise ratio (SNR), radius, and orbital period from the TOI catalog and, for planets in stellar binaries, we correct the radius for dilution by the companion. By applying a scaling relation to the measured transit SNR of each TOI in our sample, we determine the detectability of transits in each TOI as a function of both planet radius and orbital period. When applied to the full sample, this procedure elucidates the sensitivity of TESS to transiting planets as a function of binarity, host-star spectral type, planet radius, and planet orbital period. These sensitivity grids quantify the bias against the detection of small planets in unresolved binaries by TESS and show that there is a particularly low sensitivity to planets transiting secondary stars in unresolved binaries, especially as the magnitude difference between the stars increases. These sensitivity grids are available for download to facilitate their use in other studies.

\end{abstract}

% ----------------------------------------------------------------------
\section{Introduction}

Stellar multiplicity greatly complicates the interpretation of observational data of a transiting planetary system. The presence of a stellar companion can bias the inferred planetary radius by diluting a transit signal and making a planet seem smaller than it actually is \citep{ciardi15}, and in extreme cases, the dilution can conceal planets in binaries \citep{lester21}. Furthermore, it can be difficult to ascertain which of the stars in a multiple-star system is being transited by the planet \citep{payne18, lester22}. If the stars have different radii and luminosities, the uncertain identity of the host star leads to divergent estimates of the planetary radius \citep{hirsch17}. If ignored, the presence of unresolved binaries can introduce complex biases to studies of exoplanet occurrence rates \citep{bouma}. The issue of blending remains serious even if high-resolution imaging shows a planet host star to be single; the coarse pixel scale of TESS (21\arcsec) means that blending is pervasive out to $\sim1$ arcminute, and small errors in deblending techniques can propagate into significant systematic errors on the inferred planet radius and density \citep{han2025}. Moreover, the stellar parameters in the TESS Input Catalog \citep[TIC; ][]{stassun18, stassun19}, which are often used for target selection and planet characterization, explicitly presume that each star is single, so the existence of a previously unknown stellar companion can make the TIC parameters unreliable, a problem that can also affect spectroscopically derived stellar parameters of unresolved binaries \citep{furlan20}.

The issue of multiplicity is relevant in many different areas of exoplanetary science, including exoplanet occurrence-rate studies. A number of such studies have relied on a grid of planet radius and orbital period, estimating the completeness in each bin by considering only the subsample of stars whose light curve quality would allow such a planet to be detected \citep[e.g.,][]{howard12} or by performing injection-recovery tests to determine completeness \citep[e.g.,][]{dressing15}. However, both methods rely on knowing if the host stars are single or multiple in order to account for the dilution from a companion. In practice, this consideration is often not taken into account because high-resolution follow-up observations are needed to authoritatively search for evidence of a companion too close for Gaia to detect. Although it is possible to use the Gaia renormalized unit weight error statistic (RUWE) to identify possible unresolved binaries \citep{belokurov20} and to remove these stars in order to obtain a sample of single stars \citep[e.g.,][]{vach24}, this technique will only remove a subset of binaries \citep{matson25}. As a result, occurrence-rate studies that presume all planetary hosts to be single stars are biased by the presence of unidentified binary stars in ways discussed by \citet{bouma} and \citet{bergsten}. Thus, there is a need to determine how the presence of an unknown stellar companion influences the ability of TESS to detect a transiting exoplanet as a function of both the planet's radius and its orbital period.

The ongoing identification of unresolved stellar companions has also made possible the realization that planetary systems in stellar binaries are of great intrinsic astrophysical interest in their own right. Planet formation in binaries is expected to be heavily influenced by the presence of a stellar companion, leading to planetary systems that are quite different from those with single-star hosts. For example, \citet{ziegler20, ziegler21} and \citet{lester21} have shown that planet formation is suppressed in close ($<$ 100~au) binaries in TESS Objects of Interests (TOIs), and \citet{sullivan26} found that the radius valley between super-Earths and sub-Neptunes, which is a well-known feature for single-star hosts, appears to be absent in binaries.

Although there is a general understanding that an unresolved binary companion will compromise the sensitivity of TESS to small planets \citep{ciardi15}, there is still a need to quantify the effects of this diminished sensitivity. \citet{ziegler21} present one of the most detailed explorations of this issue. They injected simulated planets into the light curves of stars observed by TESS and calculated whether the resulting transit would result in SNR $>7.1$; if so, they then added a stellar companion to assess whether the dilution would obscure the transit signal. They found that $\sim$7\% of the transits detected around single-star hosts would become undetectable in the presence of a blended companion. They further concluded that $\sim$40\% of the planets in their simulations would be rendered undetectable if there were a fully blended, equally bright companion star. However, they appear to have considered only the possibility of primary-host systems, wherein the planet transits the brighter of the two stars. Another relevant study by \citet{lester21} found that small (R$\lesssim 2\mathrm{R}_{\oplus}$) planets were preferentially detected around single-star TOIs, and the authors attributed this to a bias created by the dilution of the transit signal.

In this study, we use speckle observations of host stars of TOIs, in conjunction with the TESS transit signal-to-noise ratios, to determine the sensitivity of TESS to planets of varying sizes and orbital periods in both single-star and binary TOIs.

% ----------------------------------------------------------------------
\section{Sample} \label{sample}

To create our sample, we downloaded three lists from the Exoplanet Follow-up Observing Program\footnote{\url{https://exofop.ipac.caltech.edu/tess/}} (ExoFOP) website: the full list of TOIs with transiting planet candidates, a list of all TOIs for which imaging observations have been uploaded in ExoFOP, and a list of all TOIs with stellar companions. These are dynamic, community-maintained lists, and this study uses the versions from 2025 April 2.

\begin{figure}
    \includegraphics[width=\columnwidth]{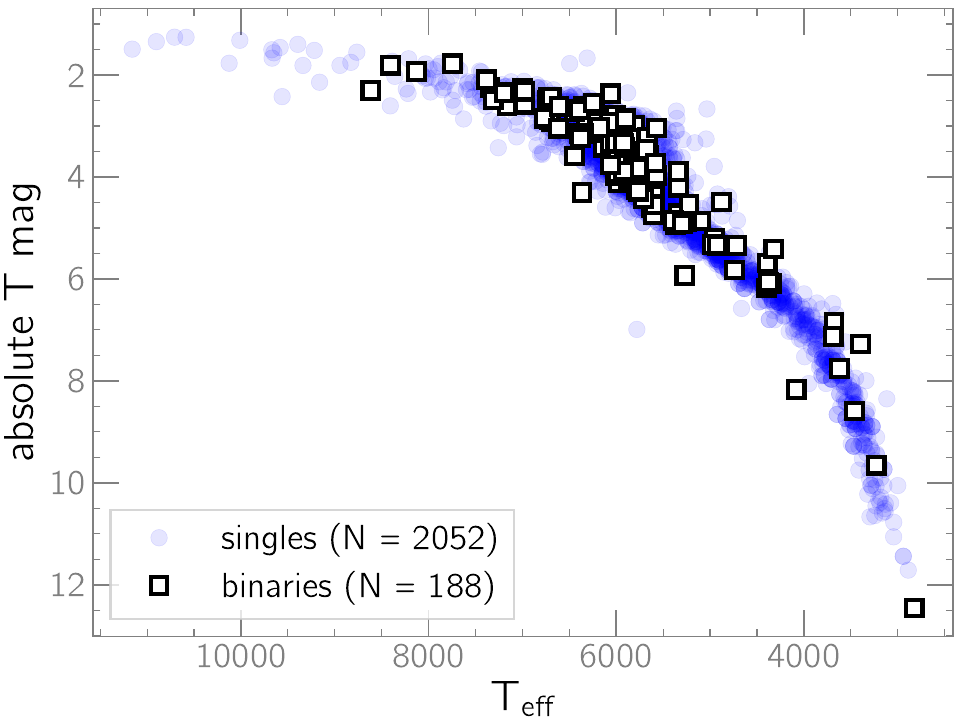}
    \caption{ Effective temperature-magnitude diagram of the single-star and binary TOI hosts in our sample.
    \label{fig:color-mag}}
\end{figure}

\begin{figure}
    \includegraphics[width=\columnwidth]{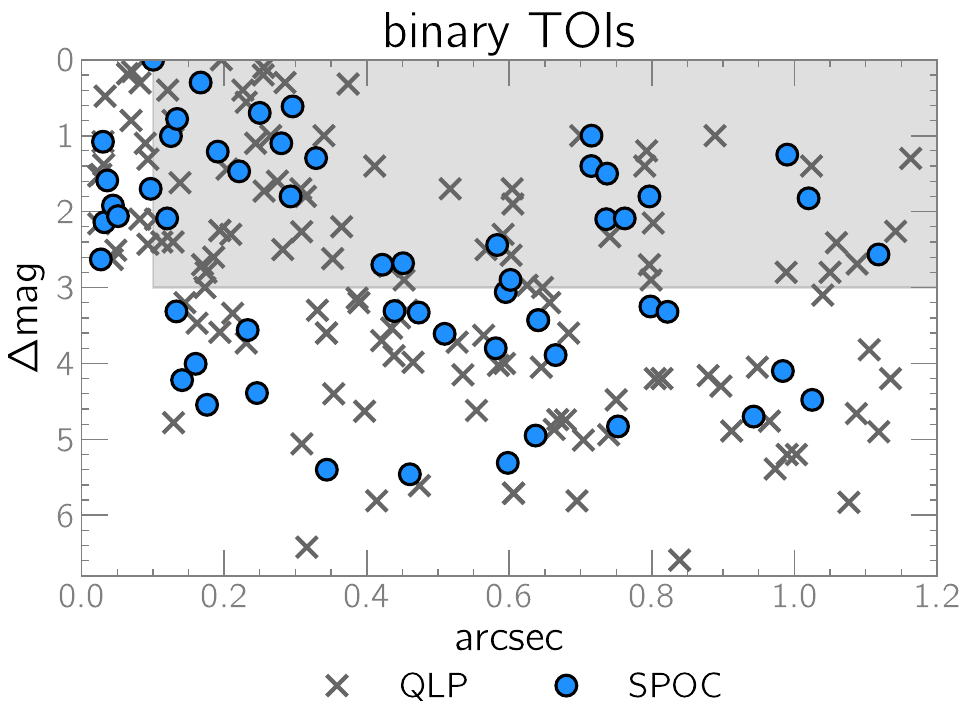}
    \caption{Separations and magnitude differences of the binaries in our sample, subdivided by the detection pipeline. There are no near-equal brightness binaries at wider separations, even though they are abundant below 0.4". The shaded region corresponds to the detection bias discussed by \citet{littlefield}. The properties of binary host stars do not correlate with the detection pipeline.
    \label{fig:binaries}
    }
\end{figure}

We kept only TOIs with main-sequence host stars and for which high-resolution speckle interferometric observations have been obtained at WIYN \citep{howell21}, Gemini \citep{lester21}, and SOAR \citep{ziegler20, ziegler21}. We selected these instruments because they have generally similar detection limits and resolutions and have been used extensively to obtain high-resolution imaging of TOIs.

ExoFOP does not formally state whether a star has been ascertained to be single, so we classified a TOI as single if it (1) has been vetted with the aforementioned speckle imagers and (2) is not in the list of TOIs with stellar companions.

Identifying and characterizing the TOI binaries in ExoFOP requires an extra step because the database often lists multiple observations---obtained with various instruments at different times---of the same stellar companion(s) without explicitly identifying which companion was observed. To arrive at a list of unique stellar companions, one must take the companion's position angle, the binary angular separation, and the binary magnitude difference (\delm) at various wavelengths and associate each measurement with a stellar companion, allowing for the possibility that there could be more than one companion. To
accomplish this, we used the Python implementation of the DBSCAN clustering algorithm in {\tt scikit-learn} \citep{pedregosa2011scikit} to group the detected companions of each TOI on the basis of their angular separation and position angle. The position angle occasionally has a 180$^{\circ}$ ambiguity in speckle observations, so to prevent this issue from resulting in a spurious number of companions, we supplied the clustering algorithm with the position angle modulo 180$^{\circ}$.

Binaries lacking valid \delm\ measurements at either 832~nm or 879~nm were excluded,\footnote{The red cameras in `Alopeke, Zorro, and NESSI typically use a filter centered on 832~nm, while HRCam uses a filter centered on 879~nm.} as were binaries with separations $>1.2$\arcsec. Above this separation, speckles typically become decorrelated, resulting in potentially unreliable measurements of the angular separation and \delm. Stellar companions with angular separations of $\leq1.2$\arcsec\ are more likely to be gravitationally bound \citep{horch14, hirsch17, matson18}. Still, to make a preliminary assessment as to whether the companions are line-of-sight companions, we performed a cone search in Gaia~DR3, with a radius of 15\arcmin\ centered on each binary TOI. We then measured the number of stars of comparable brightness to the detected companion and computed the probability of such a star randomly being as close to the TOI as the observed companion star. In all instances, the probability of a line-of-sight companion was very small (no higher than 0.5\%, and often under 0.1\%). However, the question of whether the secondaries are gravitationally bound is not particularly critical to the analysis in this study, which is concerned with the consequences of \textit{unresolved} companions that can be detected only with high-resolution imaging, regardless of whether the stars are gravitationally bound.

Next, we applied several cuts based on criteria in the full TOI catalog downloaded from ExoFOP. We limited our sample to TOIs whose TESS Follow-up Observing Program Working Group disposition is `CP' (confirmed planet), `KP' (known planet), or `PC' (planetary candidate), and we further required that planets have estimated radii in the TOI catalog of $0<\mathrm{R}_{\oplus}<20$ and periods of $0<\mathrm{P[d]}<20$. We also excluded seven systems with an SNR of exactly 1000.0 because they are actually low-SNR systems whose SNR was hard-coded in order to ensure that they would be processed.\footnote{https://tess.mit.edu/toi-releases/toi-release-notes/}

Nineteen TOIs fulfilled the selection criteria but had multiple companions, and we limited our analysis to TOIs that had no more than one speckle-detected companion.

Our final sample contains 2052 single hosts (consisting of 310 CPs, 223 KPs, and 1519 PCs) and 188 binary hosts (consisting of 10 CPs, 10 KPs, and 168 PCs). Fig.~\ref{fig:color-mag} plots the absolute magnitude as a function of stellar effective temperature for both the single and binary subsamples. Table~\ref{table} identifies the systems in our sample and provides their relevant properties.

\begin{deluxetable*}{ccccccccc}
\tablecaption{List of TOIs in sample}

\tablehead{\colhead{TOI} & \colhead{SNR} & \colhead{planet period (d)} & \colhead{planet radius (R$_{\oplus}$)} & \colhead{SpT} & \colhead{binary separation (arcsec)} & \colhead{\delm} & \colhead{Xr$_1$} & \colhead{Xr$_2$}}
\startdata
159.01 & 297.0 & 3.76 & 17.3 & F & 0.645 & 4.05 & 1.01 & 2.59 \\
161.01 & 45.0 & 2.77 & 4.6 & G & 0.196 & 0.0 & 1.41 & 1.41 \\
172.01 & 14.0 & 9.48 & 9.8 & G & 1.118 & 4.9 & 1.01 & 3.38 \\
194.01 & 203.0 & 4.90 & 16.0 & G & 0.265 & 1.0 & 1.18 & 1.39 \\
246.01 & 157.2 & 7.87 & 15.1 & K & 0.422 & 2.7 & 1.04 & 1.98 \\
272.01 & 107.8 & 3.31 & 12.4 & K & 0.032 & 2.14 & 1.07 & 2.04 \\
287.01 & 42.4 & 3.78 & 17.7 & G & 0.141 & 4.22 & 1.01 & 2.95 \\
295.01 & 65.3 & 3.40 & 14.4 & G & 0.984 & 4.1 & 1.01 & 2.84 \\
325.01 & 20.0 & 4.37 & 5.9 & K & 0.585 & 4.025 & 1.01 & 2.9 \\
364.01 & 15.0 & 0.49 & 3.3 & F & 0.374 & 0.32 & 1.32 & 1.38
\enddata
\tablecomments{A sample of the full table is provided here to illustrate its format. The full table is available for download in machine-readable format. The final four columns will be empty for all stars that have been verified to be single. $Xr_1$ and $Xr_2$ are the planetary radius-correction factors defined in Eqn.~\ref{eqn:X_r}. }
\label{table}
\end{deluxetable*}

We also show the properties of the detected binaries in Fig.~\ref{fig:binaries}. Oddly, there are no detected binaries with \delm$<1$ and angular separations between $0.4-1.2$~arcsec (Fig.~\ref{fig:binaries}). One possible explanation for this deficit is some variation of the bias identified by \citet{littlefield}, who found that binaries near this parameter space tend to have incomplete stellar parameters in the TESS Input Catalog which, in turn, can lead to their exclusion from samples. However, as shown in Fig.~\ref{fig:binaries}, the parameter space of the \citet{littlefield} bias is significantly larger than the parameter space where binaries are absent in our sample.

\begin{figure}
    \centering
    \includegraphics[width=\columnwidth]{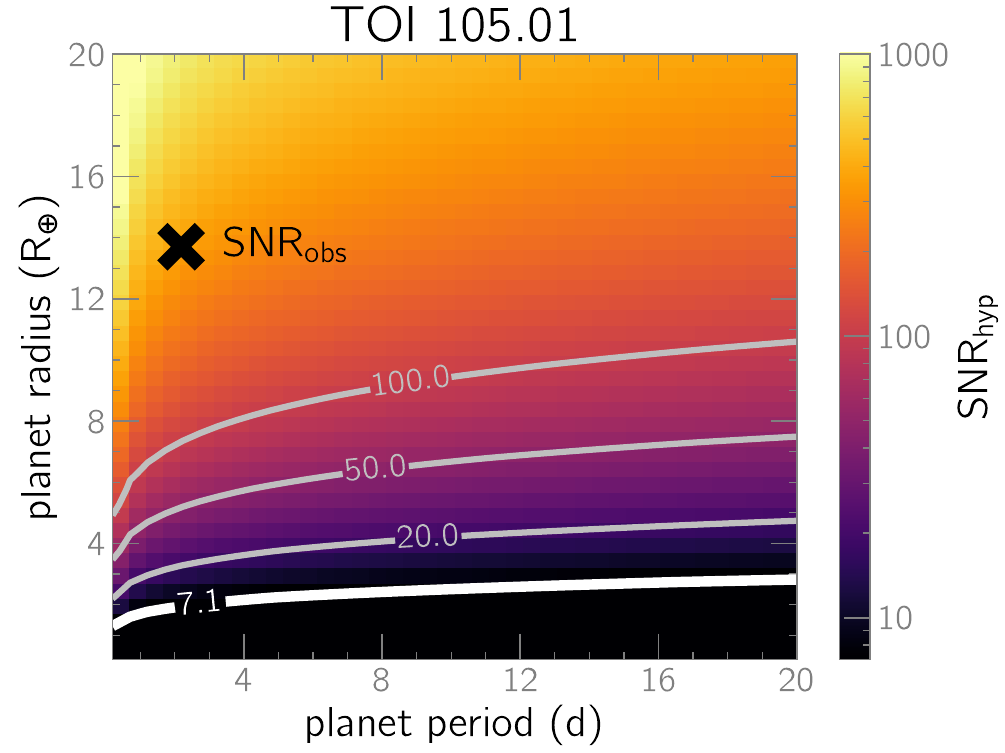}
    \caption{Representative sensitivity grid. The true period and radius of the planet are marked with an `x' and result in SNR$=349$, while the colormap and contours indicate the scaled SNR of hypothetical planets of various periods and radii.
    \label{fig:sample_sensitivity_grid}
    }
\end{figure}

\begin{figure}
\centering
\includegraphics[width=\columnwidth]{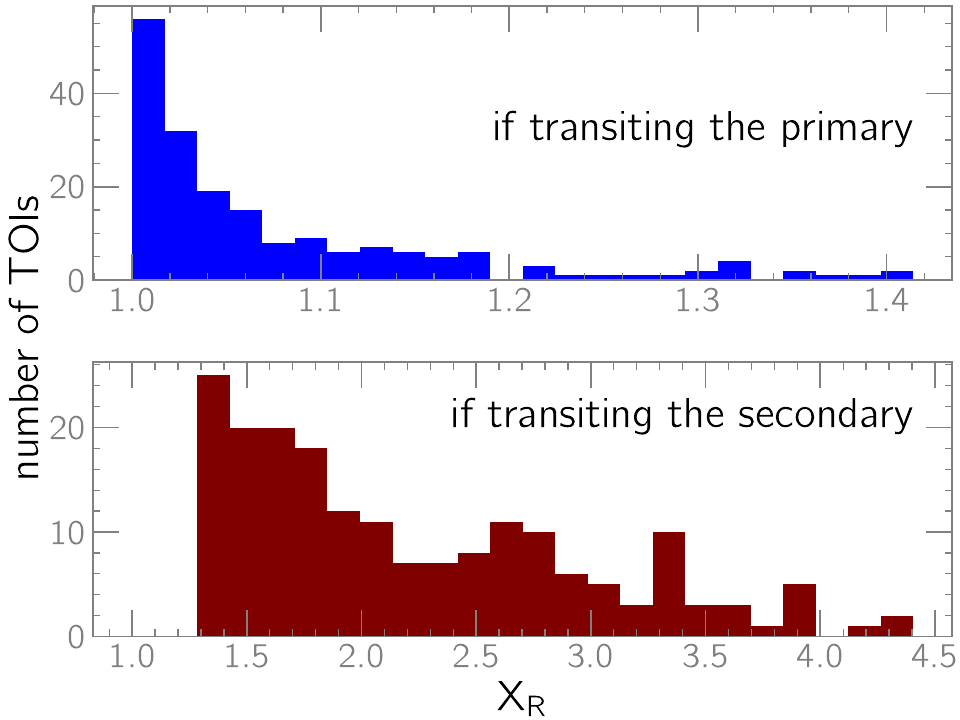}
\caption{Radius correction factors for the binary TOIs in our sample. In the top panel, the planets are all presumed to transit the primary, while in the lower panel, they are presumed to transit the secondary. Note the different $x$-axis scales in the two panels.
\label{fig:radius_correction}
}
\end{figure}

\begin{figure}
\centering
\includegraphics[width=\columnwidth]{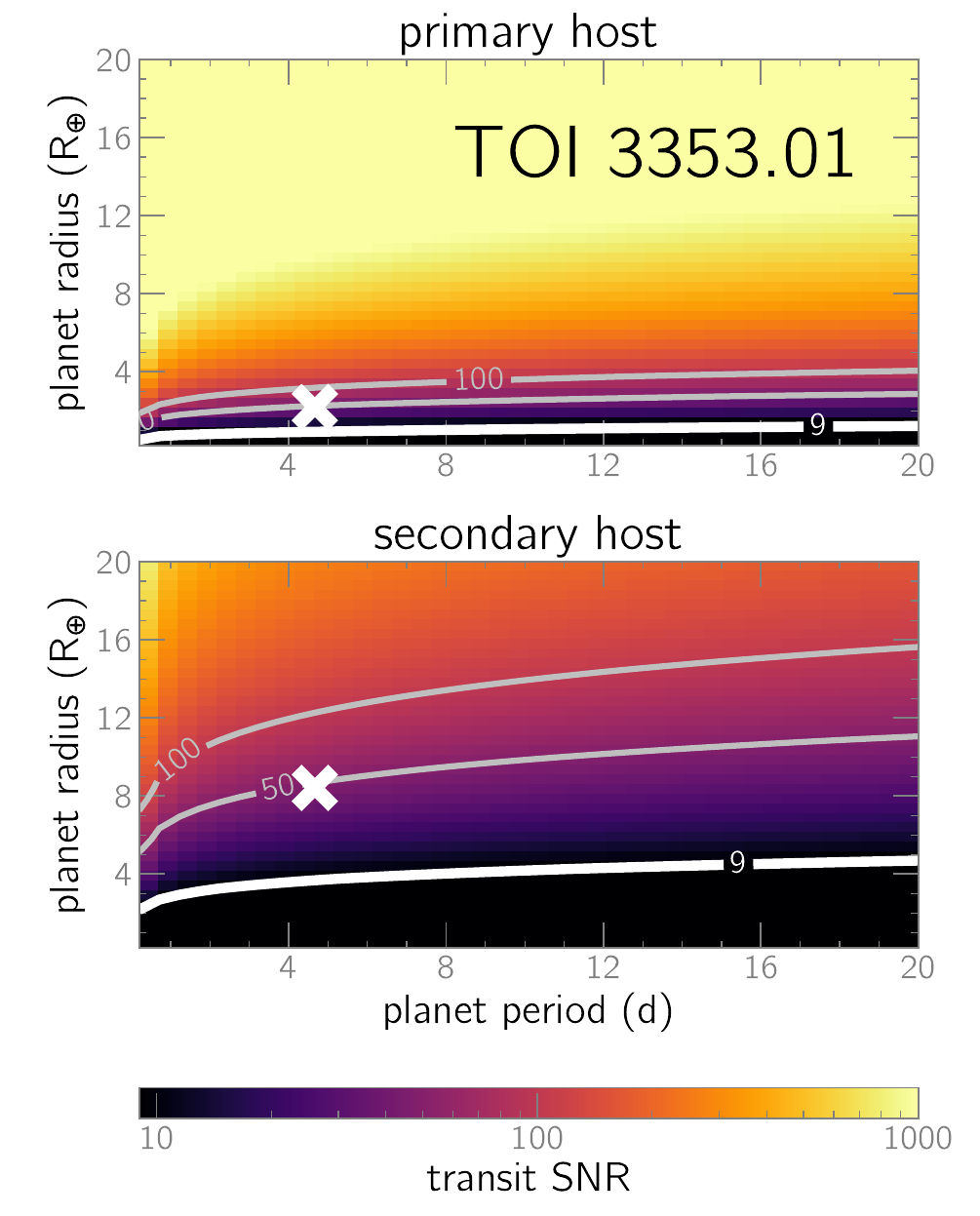}
\caption{Sensitivity curves for an illustrative binary TOI in our sample in which it is currently unknown which of the binary stars is the host of the transiting exoplanet TOI 3353.01.  In both panels, the `x' indicates the orbital period and corrected radius of the planet. An Earth-sized planet orbiting the primary or a giant planet orbiting the secondary would result in transits with the same SNR. The SNR = 9 contour is the TESS detection threshold for transits by TOIs that were detected via the QLP pipeline, as TOI 3353.01 was. While TESS would be expected to detect Earth-sized short-period exoplanets transiting the primary, it would be sensitive only to planets with $\gtrsim4$~R$_{\oplus}$ if they transit the secondary.
\label{fig:example_binary_TOI}
}
\end{figure}

% ----------------------------------------------------------------------
\section{Methodology}

\subsection{Single-star TOIs}

Exoplanet transits detected by TESS have an associated signal-to-noise ratio (SNR) that depends on the transit depth ($\delta$), the orbital period ($P$), and the photometric noise ($\sigma$), such that
\begin{equation}
SNR \propto \frac{\delta \times P^{1/3}}{\sigma},\end{equation}
where the transit depth depends on the ratio of the planetary radius to the stellar radius ($\delta \sim (R_p/R_*)^2$; \citealt{ciardi13}). The TESS SPOC pipeline uses a fixed detection threshold\footnote{In reality, transits with $SNR<7.1$ are detectable but typically standard processing does not make them TOIs; currently on the ExoFOP, there are 14 TOIs out of 7525 which have $SNR<7.1$.} of SNR$\geq7.1$ for transits, while the Quick Look Pipeline uses a nominal threshold of SNR$\geq9$ \citep{kunimoto23}. Thus, the sensitivity of TESS to planets of a particular radius and period vary from system-to-system, with a significant dependence on the radius of the star that is being transited.

Given a transiting exoplanet with radius $R_{obs}$, orbital period $P_{obs}$, and transit SNR ($SNR_{obs}$), it is possible to predict the transit SNR of hypothetical planets of different radii and orbital periods in the same system. \citet{ciardi13} showed that the SNR for each hypothetical planet ($SNR_{hyp}$) can be calculated by scaling $SNR_{obs}$ to the hypothetical radius and period, such that
\begin{equation}
SNR_{hyp} = SNR_{obs}
\times \Bigg(\frac{R_{hyp}}{R_{obs}}\Bigg)^2
\times \Bigg(\frac{P_{hyp}}{P_{obs}}\Bigg)^{-1/3}. \label{eqn:SNR}
\end{equation}

This relation assumes that the transit chord remains the same regardless of $R_{hyp}$ and $P_{hyp}$. Changing the planet's orbital period will necessarily change its semimajor axis and thus its impact parameter, defined as $b = a\cos (i)/R_*$ \citep[e.g.,][]{winn}, unless $b=0$ (or, equivalently, $i$ is exactly $90^{\circ}$). Thus, Equation~\ref{eqn:SNR} implicitly requires that the hypothetical planets have slightly different orbital inclinations compared to the actual planet. Fig.~\ref{fig:sample_sensitivity_grid} shows a representative example of a TOI for which $SNR_{hyp}$ has been calculated for periods between $1-20$~d and radii from $1-20$~R$_{\oplus}$.

%TOIs with just a single sector of data might not be suitable for Eq.~\ref{eqn:SNR} because as the orbital period increases, there is a reduced likelihood of detecting multiple transits within that single sector of data. And without multiple transits, a star would not normally be identified as the host of a transiting exoplanet. To illustrate this complication, we use the example of a TOI whose SNR is scaled from a true value of $P_{orb} = 2$~d to a hypothetical $P_{orb} = 15$~d. In just one sector of TESS data, a planet with this longer period might be observed to transit as many as two times or as few as zero, depending on its orbital phase at the start of the TESS sector. If the transit were to coincide with the mid-sector downlink gap, the system would appear to be non-transiting. Thus, the data gaps within a single sector can conceal transits in a way that Eq.~\ref{eqn:SNR} does not contemplate. We mitigate this risk by excluding all systems with just a single sector of TESS observations.

\subsection{Binary TOIs}

It is relatively straightforward to apply Eq.~\ref{eqn:SNR} to the confirmed single-star TOIs because we can simply take the transit SNR, orbital period, and planet radius directly from the TOI catalog. The verified absence of an unresolved stellar companion means that the stellar radii (and thus the inferred planetary radii) do not require correction.

But in the case of binary TOIs, the planetary radius in the TOI catalog is incorrect due to dilution of the transit depth. It is possible to correct for the dilution by computing a correction factor $X_r$, \citep{ciardi15,ciardi17,hirsch17} as
\begin{equation}
    X_r = \frac{R_{p}(true)}{R_p(single)} = \frac{R_t}{R_1}\sqrt{\frac{F_{tot}}{F_t}},
    \label{eqn:X_r}
\end{equation}
where $R_{p}(true)$ is the true planetary radius, $R_p(single)$ is the incorrect planetary radius based on the false assumption of a single star, $R_t$ is the true radius of the star being transited, $R_1$ is the nominal radius of the star under the incorrect assumption of it being single, $F_{tot}$ is the total flux of all of the stars that contribute to the transit light curve, and $F_t$ is the flux of the star being transited.

\begin{figure*}
    \centering
    \includegraphics[width=\textwidth]{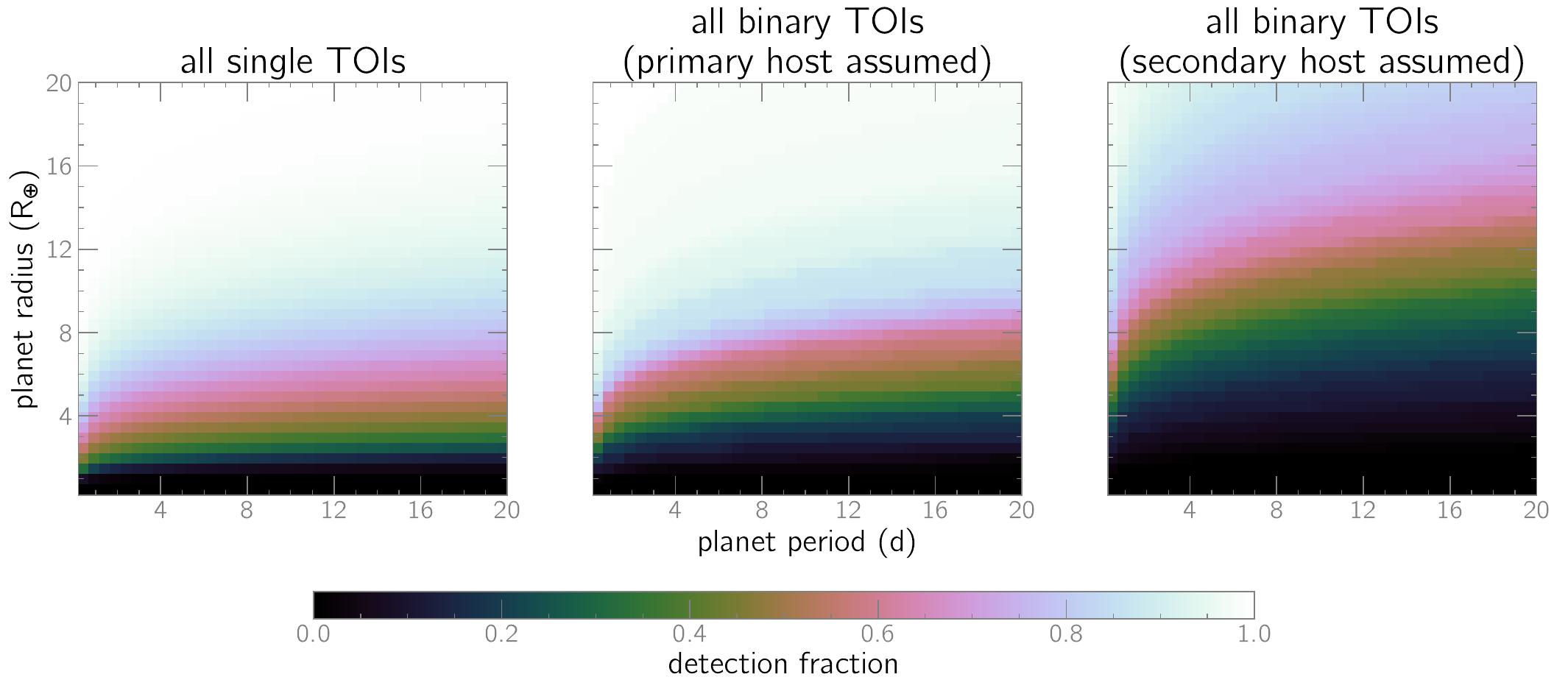}

    \caption{{\bf Left:} Sensitivity of TESS to transits in the full sample of single-star TOIs. {\bf Center:} The sensitivity for all binary stars in our sample, with all planets presumed to transit the brighter star. {\bf Right:} As with the top-center panel, except that all planets are assumed to transit the fainter star. The sensitivity of TESS to small planets around binary hosts is substantially lower than for single-star hosts, especially when the planet transits the secondary star.  \label{fig:singles_vs_binaries}   }
\end{figure*}

The difficulty with computing $X_r$ is that the identity of the exoplanet's host star is usually unknown, and due to this uncertainty, $X_r$ (and hence the dilution-corrected planetary radius) can be vastly different depending on which star is the host of the transiting exoplanet \citep{ciardi15, teske}. Figure~\ref{fig:radius_correction} illustrates this ambiguity by presenting histograms of the estimates of $X_r$ for the binary TOIs in our sample; one assumes that all planets transit the primary while the other assumes that they all transit the secondary. These are both extreme scenarios, and realistically, we would expect there to be a mix of primary- and secondary-host systems \citep{hirsch17, payne18, lester22}. Since the companion star's radius cannot simply be estimated from the TIC, we used the speckle \delm, in conjunction with the \citet{pecaut13} sequence, to estimate it when calculating $X_r$ for a secondary-hosted planet.

Because of this systematic uncertainty in $X_r$, we computed two sensitivity grids for each binary TOI, corresponding to the different values of $X_r$. An illustrative pair of sensitivity grids is shown in Figure~\ref{fig:example_binary_TOI}, demonstrating how dramatically the detection threshold can vary depending on whether the planet orbits the primary or the secondary.

\subsection{Estimating sensitivity from SNR scaling}

When sensitivity grids, such as the one in Fig.~\ref{fig:sample_sensitivity_grid}, are available for a large number of systems, it is straightforward to determine the fraction of those sensitivity grids in which a hypothetical planet of a given radius and orbital period would have fulfilled the detection criterion of the pipeline that detected the transiting exoplanet. For SPOC-detected planets, this threshold is $SNR_{hyp}\geq7.1$, while for QLP-detected planets, it is $SNR_{hyp}\geq9$. Fig.~\ref{fig:singles_vs_binaries} shows the detection fractions for our single-star and binary subsamples; in the following sections, we undertake a detailed breakdown of the sensitivity and how it depends on such factors as the binary \delm\ and the host-star spectral type.

\begin{figure*}
    \includegraphics[width=\textwidth]{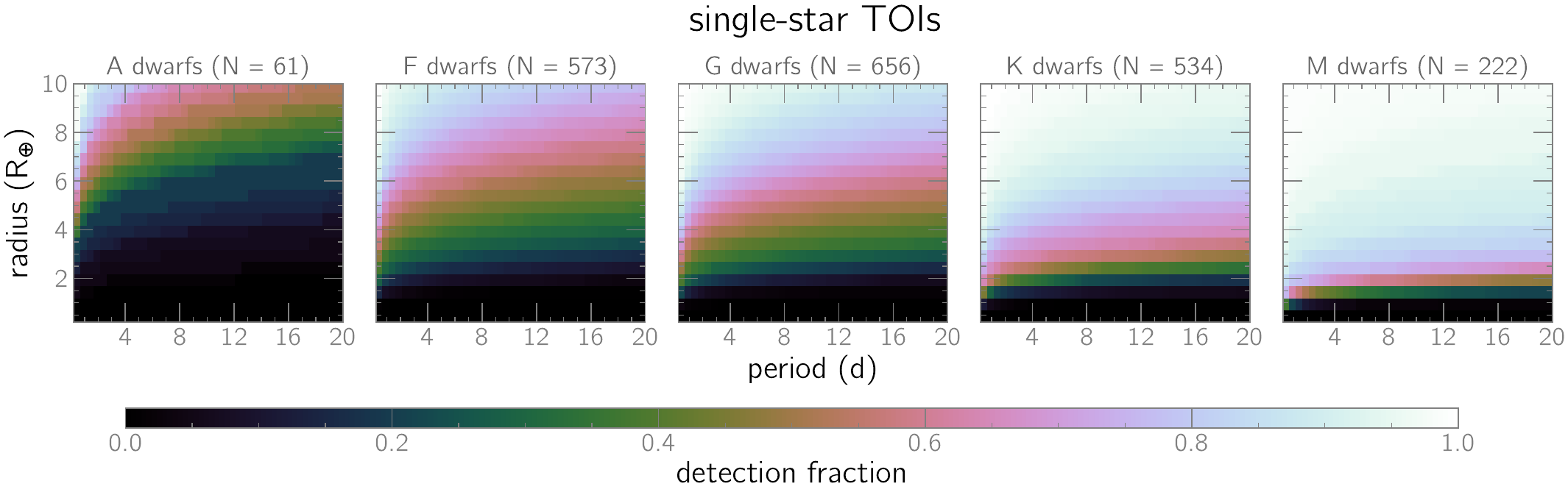}
    \caption{Sensitivity grids by spectral type for single TOIs in our sample. These data are available in machine-readable-table format to facilitate their use in other studies. Although the grids are computed up to R = 20~R$_{\oplus}$, we display the grids only to a maximum of 10~R$_{\oplus}$ in order to highlight the parameter space of the most interest.\label{fig:single_sensitivity}
    }

\end{figure*}

\section{Results}

\subsection{Single-star TOIs}

We begin by analyzing the single-star TOIs, primarily in order to establish a point of comparison for the binary TOIs.

\begin{figure*}
    \centering
    \includegraphics[width=\textwidth]{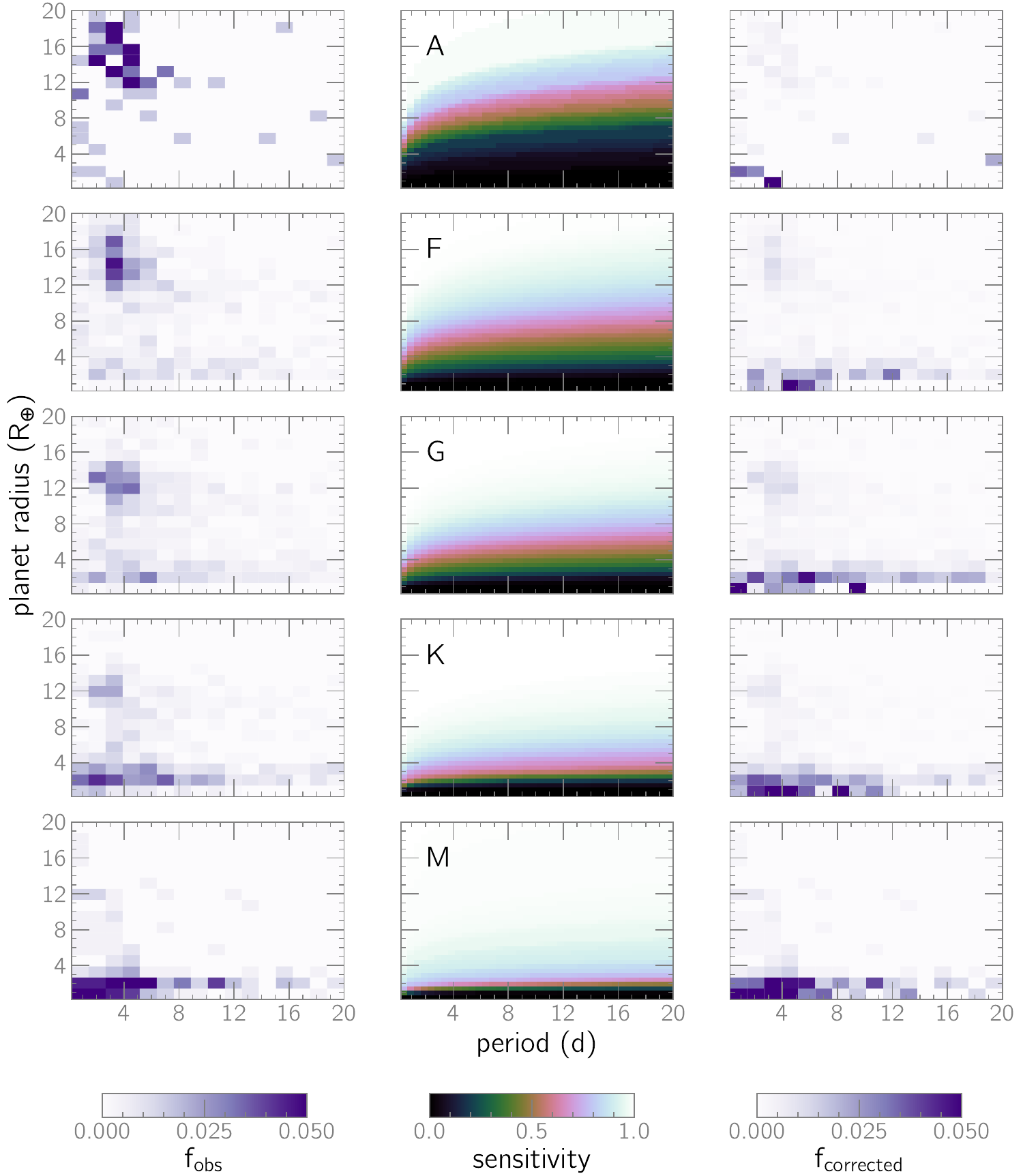}
    \caption{  {\bf Left column:} 2D histograms showing the distribution of observed transiting exoplanets around single TOI hosts of a particular spectral type. $\mathrm{f_{obs}}$ is the relative frequency of each bin within the parameter space.
    {\bf Middle column:} Sensitivity grid for that spectral type. The letter in the upper left corner indicates the spectral type for that row. {\bf Right column:} 2D histograms corrected for TESS sensitivity. $\mathrm{f_{corrected}}$ is the relative frequency of each bin after correcting for the TESS sensitivity.
    \label{fig:histograms_for_single_TOIs}
    }
\end{figure*}

We placed the single-star TOIs into different bins according to their spectral types, which we inferred from their nominal effective temperatures in the TIC and the \citet{pecaut13} sequence.
Fig.~\ref{fig:single_sensitivity} presents the resulting sensitivity grids for single-star TOIs in our sample as a function of planet radius, orbital period, and host-star spectral type. The sensitivity of TESS to small planets improves dramatically at later spectral types because the transit depth is deeper for any given planet radius. For example, the sensitivity is
$\sim90\%$ for a 5~R$_{\oplus}$ transiting planet in a 10~d orbit around an M-dwarf, but for an F-dwarf, the sensitivity for the same planet drops to about 50\%. In supplemental online material, we make these sensitivity grids available as look-up tables that can be interpolated to yield the sensitivity at any combination of radius, period, and spectral type.

These sensitivity grids can be used to estimate the period-radius distribution of transiting exoplanets if TESS were equally sensitive to all planets. Fig.~\ref{fig:histograms_for_single_TOIs} demonstrates this by representing the radius-period parameter space with a 2D histogram for spectral types A, F, G, K, and M. In the uncorrected histograms, gas giants are conspicuous for all spectral types except M and have the highest relative frequency for spectral types A, F, and G. However, after these histograms are divided by the sensitivity grid corresponding to each spectral type, the relative frequency of the gas giants is greatly reduced, while that of small planets is significantly enhanced. This trend is not nearly as apparent for spectral type A, but this can be attributed to a combination of a much smaller sample size (N = 61) than the other spectral types as well as very low sensitivity to planets smaller than 4~R$_{\oplus}$.

We emphasize that these results are distinct from exoplanet occurrence rates. While they quantify the ability of TESS to successfully detect transits by planets that do transit their host stars, our calculations do not, for example, account for the orbital geometry required for a planet to undergo transits.

% ----------------------------------------------------------------------

\subsection{Binary TOIs}

Analyzing the binary TOIs is complicated by the comparatively small number of binaries and the uncertain identity of the host. As a result, it is not feasible to reliably divide the binary hosts by inferred spectral type. Therefore, we bin the binaries by their speckle \delm, and for each of those bins, we compute two sensitivity grids: one for the scenario in which all planets transit the primary, and another in which all planets transit the secondary.

\begin{figure*}
    \includegraphics[width=\textwidth]{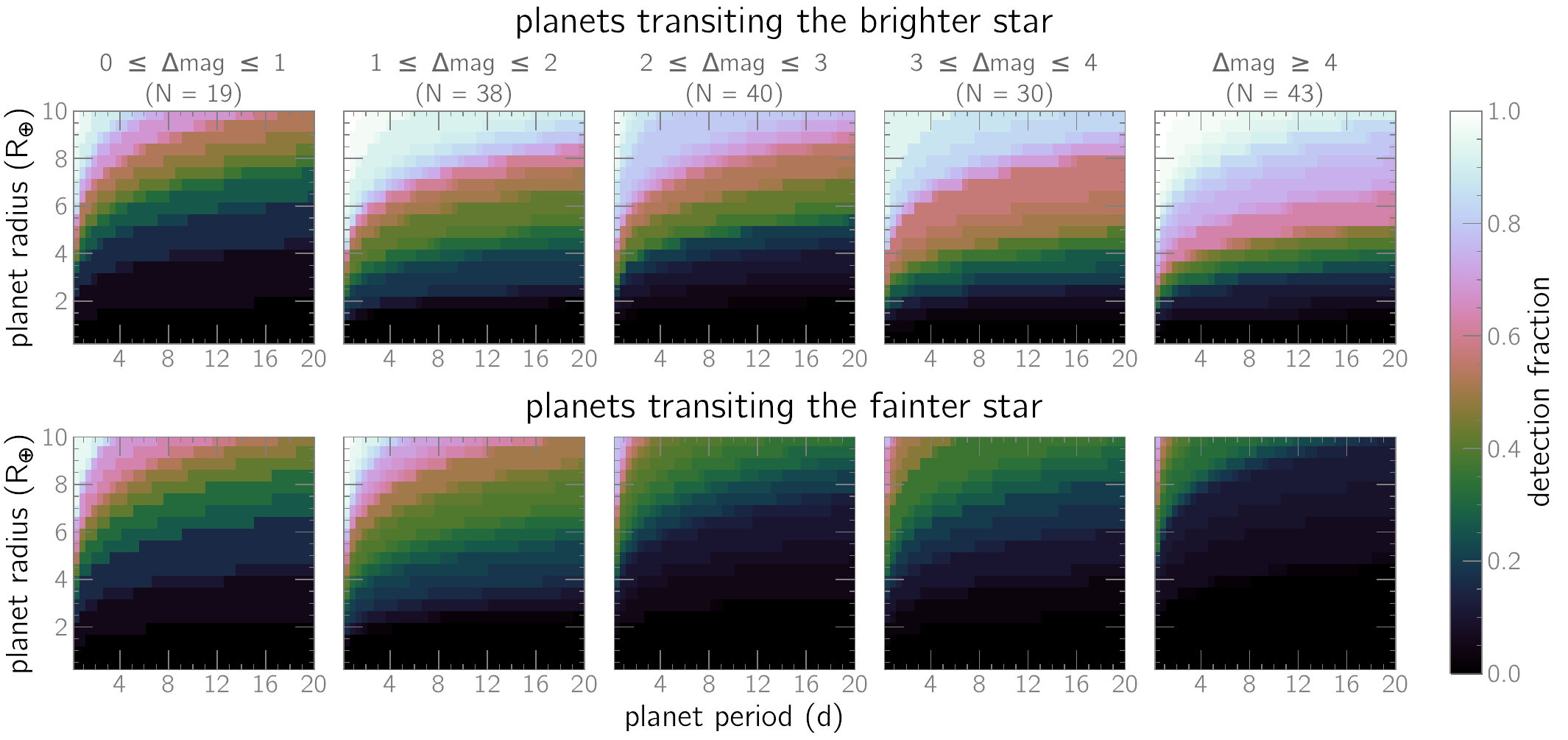}
    \caption{Effect of \delm\ on TESS sensitivity in binary TOIs. All spectral types are included in each \delm\ bin. The top row shows the sensitivity assuming that the planet orbits the primary while the bottom row assumes that the secondary is the host. The \delm\ range above each panel in the top row also describes the corresponding panel in the bottom row. At small \delm, TESS would be equally likely to detect any given planet regardless of which star it transits. However, as \delm\ increases, TESS loses its ability to detect planets transiting the fainter star.\label{fig:binary_sensitivity}
    }
\end{figure*}

We show our sensitivity grids for binary TOIs in Fig.~\ref{fig:binary_sensitivity}. For binaries with \delm$<1$, the radius-correction factors are very similar for both primary-host and secondary-host TOIs, so the sensitivity grids are almost identical. But as \delm\ increases, the radius-correction factors for primary- and secondary-host TOIs diverge; this is because the dilution becomes minimal if the planet transits a much-brighter star but extreme if it orbits a much-fainter star. For \delm$\geq4$, the sensitivity grid for primary-host TOIs is almost identical to that of a single-star TOI, with near-$100\%$ sensitivity to close-in gas giants. In contrast, for secondary-host TOIs with \delm$\geq4$, only $\sim$half of close-in gas giants would be detected and only $\sim10\%$ of Neptune-sized planets would be detected. Thus, even if both stars were equally likely to host transiting exoplanets, TESS would preferentially detect planets around the brighter star.

Geometric considerations further reduce the likelihood of detecting transiting exoplanets around the secondary. For exoplanets that have circular orbits and that are much smaller than their host star, \citet{winn} gives an approximate transit probability of $p = R_{*} / a$. \citet{bouma} point out that the stellar radius $R_*$ has a much stronger dependence on the star's mass (often approximated as $R_*\propto M_*$ for main-sequence stars) than does the semimajor axis ($a \propto M^{1/3}_*$). As a result, the secondary's comparatively small radius reduces the transit probability more than the diminished $a$ boosts it. These simple geometric considerations are not taken into account in Fig.~\ref{fig:binary_sensitivity}, so the prospects of detecting transiting exoplanets around the secondary in high-mass-ratio systems are even more dire than suggested by the sensitivity plots.

This bias against detecting secondary-host planets adds some useful context to \citet{lester22}, who examined multi-star TOI hosts in an effort to determine which star was the host of the transiting exoplanet. They determined that the primary star was the most likely host in 16 out of 23 systems in their sample; of the remaining systems, two were likely secondary hosts while the rest were indeterminate.\footnote{However, \citet{lester22} pointed out that one of the suspected secondary hosts was questionable due to (1) large uncertainties and (2) an implied radius-correction factor that would result in an unphysically low planet density.} \citet{lester22} found that the primary-host probability is slightly lower in equal-brightness binaries and becomes much higher with increased \delm. Both findings are fully consistent with the relatively low detection rate for secondary-transiting exoplanets in Figure~\ref{fig:binary_sensitivity}, as well as the aforementioned geometric considerations.

These results are also relevant to efforts to estimate planet occurrence rates in samples that include TOI host stars with blended binary companions. Binary stars are common, and so are TOIs that transit a star in a binary system \citep{ziegler20, lester21}, but in many instances, it is not possible to detect the stellar companion(s) without obtaining high-resolution imaging. This is problematic for occurrence-rate studies, which require accurate knowledge of the number of planets as well as the number of stars in a sample. The presence of unrecognized stellar companions can skew both numbers. Moreover, binaries cannot always be identified using the currently available Gaia releases; some unresolved binaries have deceptively normal RUWE \citep{matson25}, and some single stars have anomalously high RUWE \citep{ziegler20}. Thus, there is no substitute for high-resolution imaging for studies in which it is necessary to know whether a planet's host star has a stellar companion.

This reduced sensitivity will significantly complicate efforts to study the radius distribution of TOIs that orbit binary stars. The occurrence-rate study by \citet{sullivan26} for Kepler planets in binaries found fascinating differences in the radius distribution of binaries compared to single-star hosts, with an overall reduced occurrence rate compared to single-star hosts and no sign of the radius valley or the radius cliff for planets between $1-4~\mathrm{R}_{\oplus}$. A natural follow-up to their study would be to examine TOIs in binary systems for a similar tendency, but our sensitivity grids show that the TESS pipeline is unlikely to detect most such planets.

Our methodology differs from that of \citet{ziegler21} in several ways, and the two approaches are complementary. Ours allows for the possibility that the planet transits the secondary star, and it explores the poor prospects for detecting planets in such systems. Also, the binary systems simulated in \citet{ziegler21} allow for a wide range of angular separations, including those that are far in excess of our upper limit of 1.2\arcsec. Our motivation is to explore the consequences of stellar companions that will be blended in typical seeing-limited observations; in contrast, for binaries at wider separations, it becomes increasingly easy to determine which binary star is the planet host.

Our analysis is concerned primarily with assessing the reduced likelihood of detecting a planet, not with ascertaining which star in a binary TOI hosts a transiting planet. Other studies have developed techniques for attempting to make this determination. Sometimes, the identity of the host star can be accomplished by inferring the average density of the planet \citep{payne18, lester22} or by using the amplitude of the radial-velocity variations of a system to rule out one of the possible host stars. For example, \citet{rodriguez19} ruled out the possibility that TOI-172b orbits the secondary star because at the observed \delm, the companion could not produce the observed amplitude of the radial-velocity curve. Likewise, in some systems, the secondary is so faint that even if it were fully eclipsed, it could not produce the observed transit depth \citep[e.g., TOI-1135][]{mallorquin24}.

% ----------------------------------------------------------------------
\section{Conclusions}

    The multiplicity of TOI host stars has a significant effect on the detectability of exoplanets, underscoring the value of high-resolution imaging in the characterization of these systems. Unsurprisingly, if a host star is confirmed to be single, TESS is better able to detect small planets in comparison to binary TOIs, and this sensitivity improves with later spectral types.

    The situation is more complicated in binary host stars, where the ability of TESS to detect transits depends strongly on (1) the magnitude difference of the binary and (2) which of the stars hosts the exoplanet. As the magnitude difference between the two stellar components increases, TESS will become increasingly insensitive to planets around the fainter star. As a result, we would expect TESS to preferentially detect planets around the brighter star even if both stars were equally likely to host a transiting planet of a particular size and radius.

    Our analysis also suggests that there could be a population of transiting exoplanets around secondary stars in binary TOIs and that TESS will struggle to detect transits of these planets. Neptune-size  planets (and smaller) in particular have a very low probability of detection if they orbit the secondary. On one hand, this provides a justification for assuming that the primary star is the host in the absence of additional evidence, as suggested by \citet{lester22}. But it also means that the planetary content of binary systems remains poorly known in comparison to single-star TOIs.

    We quantify these effects with look-up tables, available as supplemental online material, so that future studies may correct for the sensitivity in the manner done here.

% ----------------------------------------------------------------------
\begin{acknowledgements}

We thank Jessie Christiansen for helpful comments on an earlier version of this manuscript.

KVL was supported by an appointment to the NASA Postdoctoral Program at the NASA Ames Research Center, administered by Oak Ridge Associated Universities under contract with NASA.  This work also made use of the Exoplanet Follow-up Observation Program (ExoFOP; DOI: 10.26134/ExoFOP5) website provided by the NASA Exoplanet Archive, which is operated by the California Institute of Technology under contract with the National Aeronautics and Space Administration under the Exoplanet Exploration Program.

Based on observations obtained at the international Gemini Observatory, a program of NSF NOIRLab, which is managed by the Association of Universities for Research in Astronomy (AURA) under a cooperative agreement with the U.S. National Science Foundation on behalf of the Gemini Observatory partnership: the U.S. National Science Foundation (United States), National Research Council (Canada), Agencia Nacional de Investigaci\'{o}n y Desarrollo (Chile), Ministerio de Ciencia, Tecnolog\'{i}a e Innovaci\'{o}n (Argentina), Minist\'{e}rio da Ci\^{e}ncia, Tecnologia, Inova\c{c}\~{o}es e Comunica\c{c}\~{o}es (Brazil), and Korea Astronomy and Space Science Institute (Republic of Korea).

The observations in the paper made use of the High-Resolution Imaging instrument(s) `Alopeke and Zorro. `Alopeke and Zorro were funded by the NASA Exoplanet Exploration Program and built at the NASA Ames Research Center by Steve B. Howell, Nic Scott, Elliott P. Horch, and Emmett Quigley. ‘Alopeke and Zorro were mounted on the Gemini North and South telescopes of the international Gemini Observatory, a program of NSF NOIRLab, which is managed by the Association of Universities for Research in Astronomy (AURA) under a cooperative agreement with the U.S. National Science Foundation. on behalf of the Gemini partnership: the U.S. National Science Foundation (United States), National Research Council (Canada), Agencia Nacional de Investigación y Desarrollo (Chile), Ministerio de Ciencia, Tecnología e Innovación (Argentina), Ministério da Ciência, Tecnologia, Inovações e Comunicações (Brazil), and Korea Astronomy and Space Science Institute (Republic of Korea).

This work was enabled by observations made from the Gemini North telescope. The scientific community is honored to have the opportunity to conduct astronomical research on Maunakea in Hawai‘i. We recognize and acknowledge the very significant cultural role and reverence of Maunakea to the Native Hawaiian community.

\end{acknowledgements}

\facilities{TESS}

% ----------------------------------------------------------------------
% REFERENCES
\bibliography{bib.bib}
\end{document}